\documentclass[journal]{IEEEtran}
\usepackage{graphicx}
\usepackage{refstyle}
\usepackage{amsmath}
\usepackage[flushleft]{threeparttable}

\usepackage[english]{babel}
\usepackage[dvipsnames,svgnames,x11names]{xcolor}
\usepackage[markup=underlined]{changes}

\definechangesauthor[color=BrickRed]{I}
\definechangesauthor[color=NavyBlue]{JH}

\usepackage{todonotes}
\makeatletter
\setremarkmarkup{\todo[color=Changes@Color#1!20,size=\scriptsize]{#1: #2}}
\makeatother

\begin{document}
\title{Implementation of Chua's chaotic oscillator with an HP memristor}

\author{Muratkhan Abdirash,
 Irina Dolzhikova,~\IEEEmembership{ Student Member,~IEEE,}
Alex Pappachen James,~\IEEEmembership{Senior Member,~IEEE,}
          \\
        Electrical and Computer Engineering Department, Nazarbayev University, Astana, Kazakhstan}
\maketitle

\begin{abstract}
This paper proposes an innovative chaotic circuit based on Chua's oscillator. It combines traditional realization of a non-linear resistor in Chua's chaotic oscillator with a promising memristive circuitry. This mixed implementation connects old research works that were focused on diodes with relatively new research papers that are, now, concentrated on memristors. As a result, more reliable chaotic circuit with an HP memristor is obtained that could be  used as a source of randomness. Dynamic behavior of the circuit is studied by obtaining fft analysis, different chaotic attractors and Lyapunov exponent spectrum. The results show that the addition of a memristor enhances chaotic behavior of the circuit while maintaining the same power dissipation.
\end{abstract}


\section{Introduction}\label{intro}
Memristor as a concept was introduced in 1971 by Leon Chua. He entitled memristor as "the missing fourth element", considering it as one of the basic circuit components along with a resistor, capacitor and an inductor. It was stated that a memristor would link charge with flux, completing the symmetrical dependence between the four fundamental variables of any electrical circuit: current, voltage, charge and flux \cite{chua}. 

However, the existence of a memristor was questionable up until 2008. Only at that time, laboratories at Hewlett-Packard (HP) were finally able to experimentally demonstrate a working memristor \cite{strukov}. 

After this discovery, memristors have become one of the most heavily researched topics across the world. This is, mainly, because a memristor has two main characteristics that make it extremely attractive for scholars. First, its ability to function as a memory device, which potentially can revolutionize IC industry \cite{fourth,bookJames, irmanova2017multi, irmanova2018neuron}. Second, its ability to function as a non-linear memristor. This property would be the main focus of the paper - utilizing memristor's non-linear behavior to create chaotic circuits. 

Chaotic circuits containing memristors are not something new; there have already been plenty of research on this topic. For example, in source \cite{itoh} a memristor, modeled with quadratic I-V characteristics, was used in building Chua's chaotic circuit. A very simple chaotic oscillator was proposed in \cite{PN}, in which a memristor was implemented as an active device connected in series to just an inductor and a capacitor. While in another source \cite{Lyap}, memristor with a cubic I-V dependence model was used as a non-linear element for Chua's circuit. As it can be seen, several attempts have been made in creating a chaotic circuit with a memristor. One common thing among all these papers is that a memristor was integrated to, specifically, Chua's chaotic circuit in one way or another. This is, because, firstly, Chua's circuit is very simple and, secondly, it contains a mandatory non-linear element which is what, essentially, memristor is \cite{luigi2009chua}. 

What sets apart this paper from others is that it proposes a circuit that combines two currently most popular non-linear elements, namely an HP memristor and a diode. This combination is special due to two reasons. Firstly, it makes use of two diodes in antiparallel, which have been proven many times as the most traditional and reliable realization of a non-linear element in Chua's circuit \cite{luigi2009chua}.
Secondly, it contains an HP memristor, which is, now, recognized as one of the newest realizations of a non-linear resistor. Moreover, integrating, specifically, an HP memristor, being the first ever physically realizable memristor, adds one more layer to reliability and feasibility of the circuit. It is further strengthened by using a realistic mathematical model of an HP memristor as suggested in source \cite{HP}. The proposed circuit exhibits chaotic behavior, whose output signal can then be used to generate random number sequences.

The remainder of this paper is organized in the following way. Section II provides theoretical background about the HP memristor and Chua's chaotic circuit. The methodology is given in Section III. Section IV demonstrates the results. This is followed by discussion in Section V. Finally, Section VI concludes the paper. 

\section{Background}\label{back}

\subsection{HP Memristor}\label{mem}
\begin{figure}[!h]
	\begin{center}
		\includegraphics[scale = 0.35]{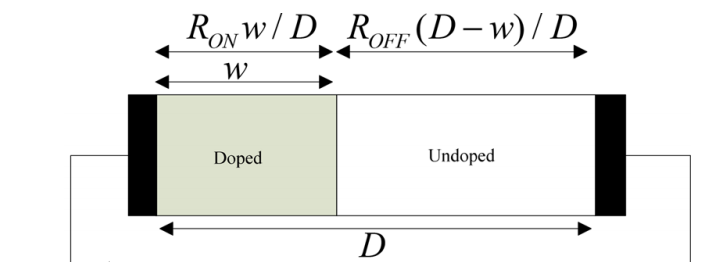}
		\caption{ Internal Structure of an HP Memristor}
		\label{fig:Mem Str}
	\end{center}
\end{figure}

An HP memristor is a thin film made of \textit{$TiO_2$}. It is doped with oxygen vacancies and bounded between two platinum plates that act as metal contacts. Its internal structure can be viewed from Fig. \ref{fig:Mem Str} as described in \cite{memr}. Here, \textit{D} is a length of the whole memristor and \textit{w} is a length of the doped region. 

Resistances of the doped and undoped regions are known as \textit{Ron} and \textit{Roff}, respectively. From this it follows that a memristor can be modeled as a series connection of two different resistances. The only complication here is that a length of the doped region changes depending on the amount and direction of the current passing through a memristor. Intuitively, when a positive current keeps on flowing through a device, the doped region expands and, the total resistance decreases. It is the opposite with a negative current resulting in the increase of a total resistance \cite{HP}. Mathematical equations describing the behavior of an HP memristor, as taken from source \cite{HP} are presented below. First, I-V equation of an HP memristor is given by Eq. \ref{eq:IV}, where $w(t)$ is a width function of memristor's doped region.

\begin{equation}\label{eq:IV}
v(t)=(R_{on}\frac{w(t)}{D}+R_{off}(1-\frac{w(t)}{D}))i(t)
\end{equation}

Secondly, \textit{w(t)} can range from 0 (when the total resistance is $R_{off}$) to $D$ (when the total resistance is $R_{on}$ and is related to current through Eq. \ref{eq:width}, where $\eta$ is a polarity and $\mu_v$ is a mobility of dopants.

\begin{equation}\label{eq:width}
w'(t)=\eta\frac{\mu_vR_{on}}{D}i(t)
\end{equation}
 
However, it was discovered that Eq. \ref{eq:IV} and Eq. \ref{eq:width} cannot fully describe the operation of an HP memristor. It happens, because  a high degree of non-linearity is observed in boundary conditions ($w\rightarrow 0$ and $w\rightarrow D$), which are far more complicated than Eq. \ref{eq:IV} and Eq. \ref{eq:width} \cite{HP}. For more comprehensive description of the model, window functions were introduced. In this paper, Biolek window is utilized since it performs better than Joglekar, but is not as complicated as other window functions \cite{HP}. Modified Eq. \ref{eq:width} with window function is represented by Eq. \ref{modif}, where Biolek window is defined by Eq. \ref{biolek}. In Eq. \ref{biolek}  $stp(i)=1$ for $i>=0$, and $stp(i)=0$ for $i<0$.

\begin{equation}\label{modif}
w'(t)=\eta\frac{\mu_vR_{on}}{D}i(t)F(\frac{w(t)}{D})i(t),
\end{equation}

\begin{equation}\label{biolek}
F_B(x,i,p)=1-(x-stp(-i))^{2p},
\end{equation}

\subsection{Chua's Chaotic Circuit}\label{chuas}

\begin{figure}[b]
	\begin{center}
		\includegraphics[scale = 0.43]{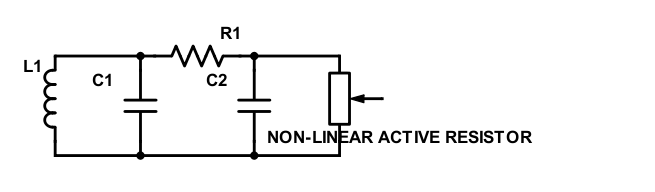}
		\caption{ Schematics of Chua's chaotic oscillator}
		\label{fig:Chua}
	\end{center}
\end{figure}

Chua’s oscillator is the simplest circuit that can generate chaos \cite{luigi2009chua}. It is demonstrated on Fig. \ref{fig:Chua} and, it is, indeed, very simple comprising of only five elements in total: two capacitors, one inductor, one resistor and a non-linear active resistor. 

Yet it is considered as a very powerful circuit that defines and sets chaotic requirements for all the other “complex” circuits out there. Moreover, Chua’s circuit is autonomous, meaning that it produces a time-varying output without a time-varying input \cite{luigi2009chua}. Having all these characteristics it establishes necessary conditions for a certain circuit to exhibit chaos it should have, at least 3 storage elements, one active local resistor and a non-linear resistor. 

Output of any circuit based on Chua's oscillator should satisfy two conditions to be recognized as chaotic \cite{luigi2009chua}: 

\begin{enumerate}\label{conditions}
\item Circuit variables (current and voltage) as measured from any node of the circuit should be chaotic and random. In other words, time plots of variables should look like a noisy signal.
\item Chaotic attractors (plot of voltage across one capacitor versus voltage across another) should be shaped as a double scroll or come close to it.
\end{enumerate}

\section{Methodology}\label{method}
\subsection{Analysis of traditional design of Chua's circuit}\label{diodes}

A lot of sources refer to diode implementation of Chua's circuit as the most popular one. For this reason, investigating how that circuit operates would be a good starting point toward the final design. It was, indeed, beneficial since the proposed circuit is, essentially, based on that classical model with diodes. Simulation files of Chua's circuit with diodes on LTspice were available on this open source \cite{spice}. 

The most important component in traditional Chua's circuit is a non-linear resistor implemented as two antiparallel diodes. This is because, ideally, memristor should be able to either replace or enhance those diodes and produce equally, if not more, random signal. To find out how exactly those diodes contribute to the overall chaotic behavior, DC sweep analysis on LTspice was performed. The circuit for DC sweep can be found in the appendix. 

\begin{figure}[b]
	\begin{center}
		\includegraphics[scale = 0.37]{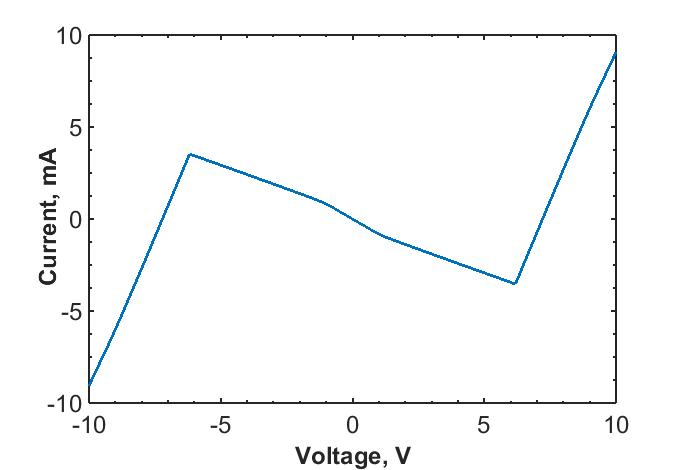}
		\caption{I-V curve of the diodes}
		\label{fig:IV}
	\end{center}
\end{figure}

After running the test circuit, I-V curve of the diodes was obtained (Fig. \ref{fig:IV}). It perfectly matches what is described in many sources including \cite{luigi2009chua}, which call this type of non-linear IV relationship as piecewise linear. The two branches with positive slope on the right and left can be ignored right now for the purpose of the analysis. The middle part of the curve in Fig. \ref{fig:IV} comprises of three line segments with different negative slopes. The negative slopes are coming from a negative impedance op amp attached in parallel to the diodes, so that in combination they function as one locally active non-linear resistor. This I-V curve demonstrates a very simple way to achieve a non-linear relationship, just by changing the slopes of a straight line. Obviously, there are other ways to achieve that - quadratic, cubic, etc. 

There is one more condition to obtain chaos - specific values of \textit{R, L} and \textit{C} parameters in the circuit. Slight changes in the values of these parameters may cause chaotic circuit to behave as a periodic oscillator \cite{luigi2009chua}.

\subsection{Integrating an HP memristor to Chua's chaotic oscillator}\label{HP}
LTspice model for an HP memristor with Biolek window was downloaded from \cite{biolek}.
The first attempt at including a memristor to the circuit was by replacing one of the diodes in antiparallel. After removing the diode, adding an HP memristor and choosing appropriate values for \textit{R, L} and \textit{C} parameters, simulation was run. However, the output chaotic signal was decaying too fast compared to the original circuit. In other words, it produced a transient chaotic signal for a relatively short period of time. That is why, it was decided to keep the both diodes.

\begin{figure}[!h]
	\begin{center}
		\includegraphics[scale = 0.4]{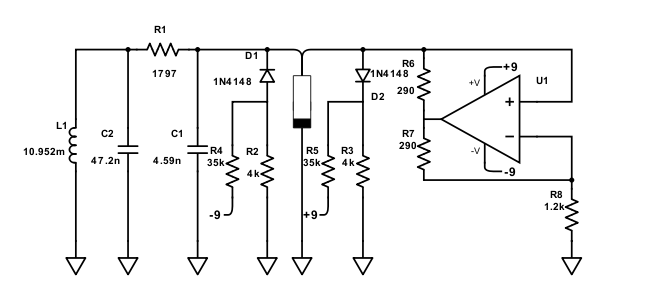}
		\caption{Schematic of the proposed circuit}
		\label{fig:schem}
	\end{center}
\end{figure}

Finally, the proposed circuit's schematic can be viewed in Fig. \ref{fig:schem}. There are three components that make the non-linear resistor for Chua's chaotic oscillator in the final design of the circuit: two diodes in antiparallel connected to an HP memristor in parallel. To confirm that the proposed circuit functions as desired, i.e. produces chaotic signals, several analysis were performed.

\begin{figure}[!h]
	\begin{center}
		\includegraphics[scale = 0.37]
        {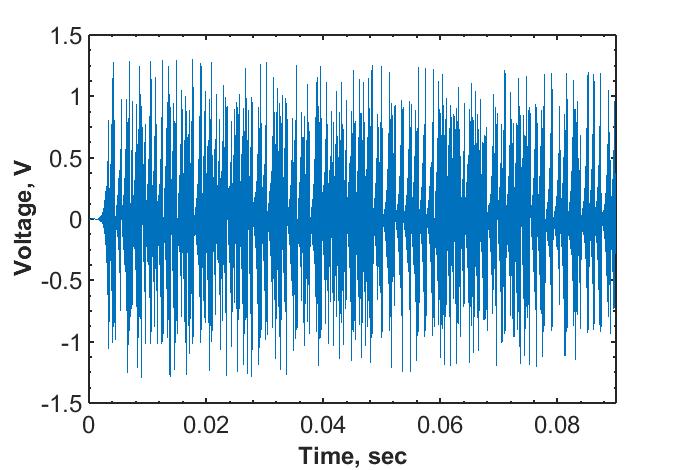}
		\caption{Random Vout of the circuit}
		\label{fig:random}
	\end{center}
\end{figure}

First of all, the output random signal of the new circuit is shown in Fig. \ref{fig:random} for 90 ms of the simulation, performed using a transient analysis in LTspice. The signal is taken as the voltage across capacitor $C2$. Visually, it is reassuring that $V_{out}$ is, indeed, random, since it behaves like a very noisy signal with no pattern whatsoever.

\begin{figure}[!h]
	\begin{center}
		\includegraphics[scale = 0.37]
       {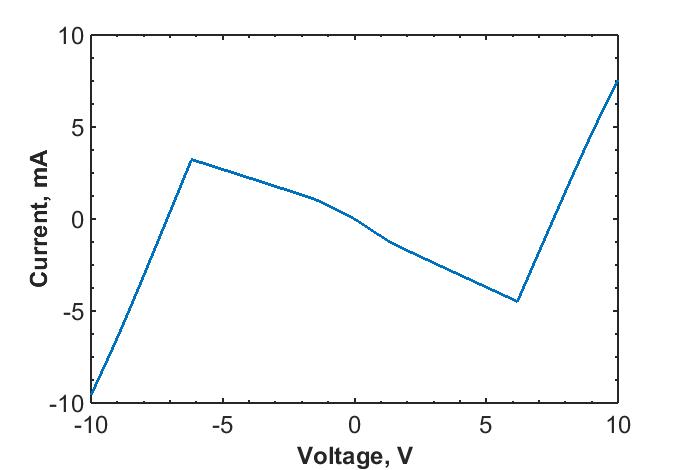}
		\caption{I-V curve of a non-linear resistor implemented with an HP memristor and a diode}
		\label{fig:IVmem}
	\end{center}
\end{figure}

Secondly, IV curve of these complex non-linear elements was obtained using DC sweep analysis in LTspice. The result is shown in Fig. \ref{fig:IVmem} and, it can be seen that an IV curve is, in fact, non-linear having different slopes and even greater number of slopes than the original circuit. This indicates how the addition of the memristor has enhanced the overall non-linearity of the circuit.  

\begin{figure}[!h]
	\begin{center}
		\includegraphics[scale = 0.37]{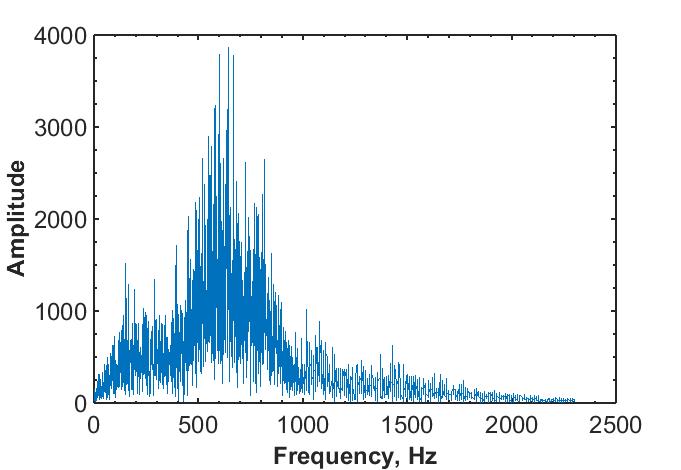}
		\caption{FFT of the output signal}
		\label{fig:FFT}
	\end{center}
\end{figure}

Thirdly, to prove the output signal's randomness or noisiness, FFT analysis was done. The results are presented in Fig. \ref{fig:FFT}. As expected, there are a lot of frequency components with different amplitudes, which is exactly what is required from a random signal.

\begin{figure}[!h]
	\begin{center}
		\includegraphics[scale = 0.37]
        {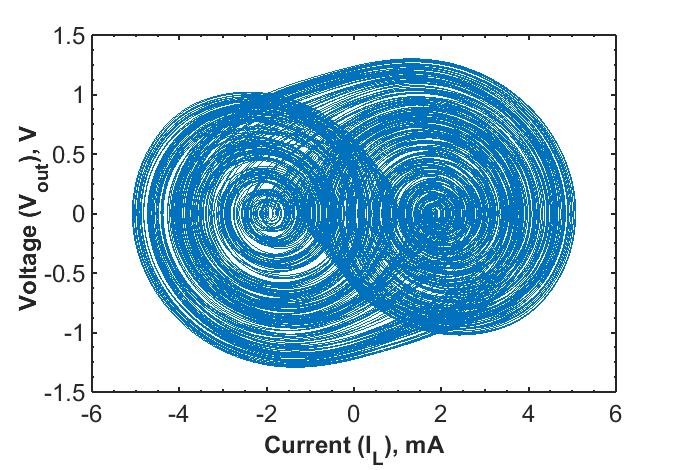}
		\caption{Chaotic attractor of I vs. V}
		\label{fig:attractor}
	\end{center}
\end{figure}

The next checking point was to plot the chaotic attractor of the system. As it was mentioned earlier, chaotic system should have an attractor shaped like a double scroll. Chaotic attractor is plotted as one circuit variable against another. In this case, voltage across capacitor $C2$ was plotted against the current in the inductor $L1$. It is shown in Fig. \ref{fig:attractor} and, it is shaped as a double scroll and satisfies the condition for chaotic system.

\begin{figure}[!h]
	\begin{center}
		\includegraphics[scale = 0.37]{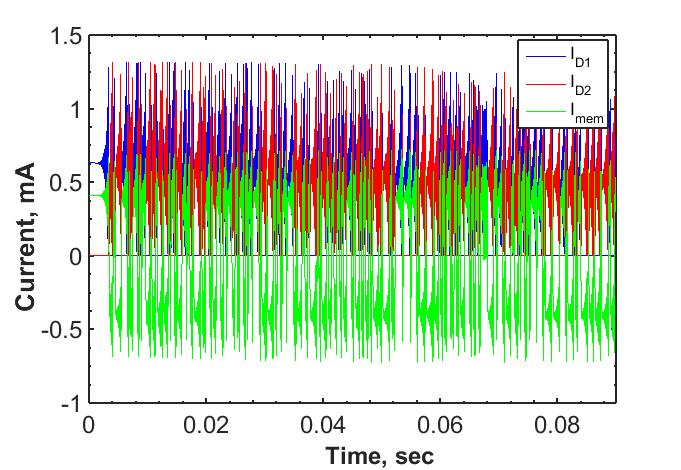}
		\caption{Currents in D2, D1 and memristor}
		\label{fig:currents}
	\end{center}
\end{figure}

Lastly, to make sure, that randomness is coming from both diodes and memristor, the currents in each component are shown in Fig. \ref{fig:currents}. It is clear, that each component is equally contributing to the randomness.

\section{Results}\label{results}
\subsection{Derivation of differential equations governing the circuit}
To study the dynamics of Chua's circuit, one must derive the differential equations governing the behavior of that circuit. Here, dynamics means the response of the circuit under changing parameter values. Obviously, parameters refer to \textit{R, L and C} values integrated to the circuit. The process of deriving those equations is described in source \cite{luigi2009chua} and all the proceeding equations are based on that.
\subsubsection{I-V Mathematical Relationship}
I-V curve of the non-linear element of the proposed circuit is presented in Fig. \ref{fig:IVmem}. As it can be seen, there are four regions with different negative slopes in the I-V graph. 

\subsubsection{Slope 1}
Point 2: (-6.1799 V, 3.2334 mA);
\par Point 1: (-1.2700 V, 0.9300 mA);
$$A=\frac{(3.2334-0.9300)*10^{-3}}{-6.1793+1.2700}=-0.4691 mS$$

\subsubsection{Slope 2}
Point 2: (-1.2700 V, 0.9300 mA);
\par Point 1: (0 V, 0 mA);
$$B=\frac{(0.9300-0)*10^{-3}}{-1.2700-0}=-0.7323 mS$$

\subsubsection{Slope 3}
Point 2: (0 V, 0 mA);
\par Point 1: (1.2300 V, -1.1500 mA);
$$C=\frac{(0+1.1500)*10^{-3}}{0-1.2300}=-0.9350 mS$$

\subsubsection{Slope 4}
Point 2: (1.2300 V, -1.1500 mA);
\par Point 1: (6.1820 V, -4.4850 mA);
$$D=\frac{(-1.1500+4.4850)*10^{-3}}{1.2300-6.1820}=-0.6735 mS$$

\subsubsection{I-V Equation}It is clear from Fig. \ref{fig:IVmem} that I-V relationship is piece-wise linear and, equation will be in accordance to that (only considering negative slopes of the graphs):
\begin{equation}
 i_R = \begin{cases} 
      Av_R-1.27(A-B), & v_R\leq -1.27 \\
      Bv_R, & -1.27\leq v_R\leq 0\\
      Cv_R, & 0\leq v_R \leq 1.23 \\
      Dv_R+1.23(C-D), & v_R \geq 1.23 
\end{cases}
\end{equation}
where $i_R$ and $v_R$ are the current in and the voltage across the non-linear element, respectively. Meanwhile, $A, B, C$ and $D$ are the slopes of the I-V curve.
\subsubsection{Differential Equations}
Following the guidelines from source \cite{luigi2009chua}: let us indicate as $v_1, v_2$ and $i_L$ the voltage across capacitor C1, the voltage across capacitor C2 and the current in the inductor L, respectively. By applying the Kirchhoff's circuit laws and and considering I-V dependence formulas for a capacitor and an inductor, the state equations can be derived as follows:
\begin{equation}
\frac{dv_1}{dt}=\frac{1}{C_1}(\frac{v_2-v_1}{R}-i_R)
\end{equation}
\begin{equation}
\frac{dv_2}{dt}=\frac{1}{C_2}(\frac{v_1-v_2}{R}+i_L)
\end{equation}
\begin{equation}
\frac{di_L}{dt}=-\frac{1}{L}v_2
\end{equation}

\subsection{Lyapunov Exponents}
To experimentally verify chaos, the Lyapunov exponents should be calculated. This type of exponents may be computed using the time series method as suggested by \cite{Lyap}. There can be several Lyapunov exponents and, as long as any of them is positive, the system is said to be chaotic \cite{Lyap}.
\par Prior to obtaining the Lyapunov exponents, all of the differential equations describing the system are scaled by $\sqrt[]{|L_1C_2|}$. It is required since the
Lyapunov exponent algorithms numerically converge for, particularly, these time factors. At lower time constant coefficients, very small step sizes are needed to solve the differential equations, which, in turn, leads to unnecessary computing complications. It does not only take a lot of time, but sometimes may result in considerable errors \cite{Lyap}.
\begin{figure}[!h]
	\begin{center}
		\includegraphics[scale = 0.37]{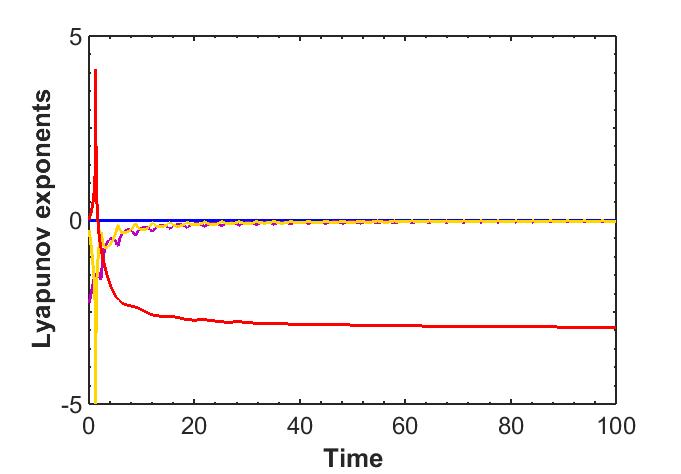}
		\caption{Dynamics of Lyapunov exponents}
		\label{fig:Lyap}
	\end{center}
\end{figure}

The result is shown in Fig. \ref{fig:Lyap}, where the red curve corresponds to the largest Lyapunov exponent. This line is above the x-axis for almost 1.7 time series simulation, meaning it is positive for just enough time. Referring back to the first paragraph in this section, the condition for chaos is satisfied.

\subsection{Power Calculations}

Since both the current in the inductor and the voltage across it are random, their product, i.e. power, is also random. To resolve this issue, average of all the instantaneous power values during 300 ms will be referred as "power absorbed by a particular element". It should be noted that the average value would still vary depending on the duration of the simulation. Very small negative power shows that inductor does not consume much of the energy and, thus, is not an issue from a power perspective. 

\begin{table}[!ht]
\centering
\renewcommand{\arraystretch}{1.15}
\caption{Comparison table for the power consumption}
\label{table}
\noindent
\begin{threeparttable}
\begin{tabular}{|c|c|c|}
\hline 
Circuit elements & Original circuit & \begin{tabular}{@{}c@{}} Memristor based \\circuit \end{tabular}
\tabularnewline
\hline
\hline
Energy storage elements &  5.6464 $mW$& 4.6099 $mW$ \tabularnewline
\hline 
Non-linear elements&  4.8894 $mW$ & 6.2303 $mW$  \tabularnewline
\hline 
Negative resistor&32.8972 $mW$  &   31.7379 $mW$  \tabularnewline
\hline 
Total power consumption& 43.4330 $mW$ &  42.5871 $mW$  \tabularnewline
\hline
\end{tabular}
\end{threeparttable}
\end{table}

Table \ref{table} demonstrates the summary of the power consumed be the proposed and original Chua's circuit.
Total power consumed by the proposed circuit during 300 ms of simulation is, then, $42.5781 mW$. As it can be seem from the results, the most power is absorbed by the negative resistor components, which is constructed using  Opamp.
Consumption by energy storage elements, not including resistor $R_8$, is so negligible that can be ignored. They do not present a problem from a power perspective for, particularly, this circuit, although they usually take up a lot of area in a chip. 

Original circuit differs from the proposed circuit in in the following way:
\begin{enumerate}
\item It has a memristor in-between the diodes to enhance the non-linearity and, hence, the overall randomness;
\item The parameter values (\textit{$C_{1}$, $C_{2}$, $L_{1}$, etc.}) differ.
\end{enumerate}


\par If the results are to be compared, the total power absorbed by the original circuit is greater by $0.8459 mW$. This means that the proposed circuit with an HP memristor is 1.95 per cent more efficient than the original circuit in terms of the power consumption. Even if it is a small difference, it can be stated that the proposed circuit outperforms the original one.

\section{Discussion}
Simulation results confirm that the proposed circuit is, indeed, giving a random output signal as expected. FFT analysis show that there are many frequency components with different amplitudes in the output signal. This randomness is, essentially, coming from transient behavior of the circuit. Double scroll chaotic attractor also confirms that the system's behavior is chaotic. However, to further study the dynamics of the circuit, Bifurcation diagrams should be constructed, which also help to determine at what exact values circuit becomes chaotic.
\section{Conclusion}
This paper has suggested one way of making a chaotic random number generator with an HP memristor. The proposed circuit was based on the classical implementation of Chua's circuit - two diodes as a non-linear element. It kept both of the diodes and integrated an HP memristor. As a result, it was able to maintain chaos and randomness exhibited by the circuit. This is what makes it special: mixing a classical implementation (diodes) of a non-linear element with a moder one (memristor). Since the circuit generates chaotic random signal, its output, potentially, can be used in applications such as image encryptions, secret communications and other various
information safeties.  
\par There are two future suggestions:
\begin{enumerate}
\item The output signal of the circuit decays as time passes and loses its randomness. This should be fixed, since ideally the circuit should remain chaotic forever;
\item Statistical tests should be conducted on the values of obtained random output signal to ensure that the circuit can be used in security applications.
\end{enumerate}


\begin{thebibliography}{1}
\bibitem{chua}
L. Chua, \emph{Memristor-The missing circuit element}, IEEE Transactions on Circuit Theory:vol.~18, num.~5 \hskip 1em plus
0.5em minus 0.4em\relax September 1971.


\bibitem{strukov}
 D.B. Strukov, G. S. Snider, D. R. Stewart, and R. S. Williams, \emph{The missing memristor found}, Nature Publishing Group, vol.~453, num.~7191, p.80, \hskip 1em plus
0.5em minus 0.4em\relax 2008.
 
\bibitem{fourth}
L. Chua, \emph{The fourth element},in Memristor Networks. Springer p.1-13, \hskip 1em plus
0.5em minus 0.4em\relax 2014.

\bibitem{bookJames}
A. James, \emph{Memristor and Memristive Neural Networks}, Intech, \hskip 1em plus
0.5em minus 0.4em\relax 2018, DOI: 10.5772/6653.



\bibitem{irmanova2017multi}
A. Irmanova and A. P. James,  \emph{Multi-level Memristive Memory with Resistive Networks}, arXiv preprint arXiv:1709.04149,  \hskip 1em plus
0.5em minus 0.4em\relax 2017.

\bibitem{irmanova2018neuron}
A. Irmanova and A. P. James,  \emph{Neuron inspired data encoding memristive multi-level memory cell}, Analog Integrated Circuits and Signal Processing, Springer, p.1-6, \hskip 1em plus
0.5em minus 0.4em\relax 2018.



\bibitem{itoh}
M. Itoh and L. Chua,\emph{Memristor oscillators}, International Journal of Bifurcation and Chaos, World Scientific, v.18, num.11, p.3183-3206, \hskip 1em plus
0.5em minus 0.4em\relax 2008. 

\bibitem{PN}
P. Jin, G. Wang, Guangyi and S. Jiang, \emph{Design of PN sequence generator based on memristor oscillator},  2015 IEEE 16th International Conference on Communication Technology (ICCT), IEEE, p.51-55, \hskip 1em plus
0.5em minus 0.4em\relax 2015.

\bibitem{Lyap}
B. Muthuswamy,  \emph{Implementing memristor based chaotic circuits},  International Journal of Bifurcation and Chaos, World Scientific, v.20, num. 05, p.1335-1350, \hskip 1em plus
0.5em minus 0.4em\relax 2010.



\bibitem{luigi2009chua}
F. Luigi, and F. Mattia, and Gabriella, M. Xibilia,  \emph{Chua's Circuit Implementations: Yesterday, Today And Tomorrow},  World Scientific, v.65, \hskip 1em plus
0.5em minus 0.4em\relax 2009.

\bibitem{HP}
A. Buscarino, and L. Fortuna, and F. Mattia Frasca and L. V. Gambuzza \emph{A chaotic circuit based on Hewlett-Packard memristor},  Chaos: An Interdisciplinary Journal of Nonlinear Science, v.22, nu. 2, 023136 \hskip 1em plus
0.5em minus 0.4em\relax 2012.


\bibitem{memr}
G. Wang, and S. Zang, and X. Wang, and F. Yuan and H. H.-C. Iu \emph{Memcapacitor model and its application in chaotic oscillator with memristor},  Chaos: An Interdisciplinary Journal of Nonlinear Science, v.27, nu. 1, 013110 \hskip 1em plus
0.5em minus 0.4em\relax 2017.


\bibitem{spice}
Jim, \emph{SIMULATING CHUA’S CIRCUIT WITH LTSPICE},  [Online] Available:url{http://www.chaotic-circuits.com} \hskip 1em plus
0.5em minus 0.4em\relax February 2018.



\bibitem{IEEEhowto:biolek}
T. Michea, \emph{nowm/memristor-models-4-all},  [Online] Available:url{https://github.com/knowm/memristor-models-4-all} \hskip 1em plus
0.5em minus 0.4em\relax February 2018.









  


\end{thebibliography}

 \end{document}